\documentclass[useAMS,usenatbib,usegraphicx]{mn2e} 

\pdfoutput=1 

\usepackage{amssymb} 
\usepackage{times} 
\usepackage{aas_macros} 
\usepackage[usenames]{color} 

\newcommand{\Kepler}{{\it Kepler} }

\newcommand{\dsct}{$\delta$~Sct }

\title[Combining WASP and \Kepler data]{Combining WASP and \textit{Kepler} data: the case of the $\delta$~Sct star KIC~7106205}

\author[D. M. Bowman et al.] 
{Dominic M. Bowman,$^1$\thanks{Email: dmbowman@uclan.ac.uk} Daniel L. Holdsworth,$^2$ and Donald W. Kurtz$^1$ \\
$^{1}$Jeremiah Horrocks Institute, University of Central Lancashire, Preston PR1 2HE, UK \\
$^{2}$Astrophysics Group, Keele University, Staffordshire, ST5 5BG, UK}
 
\date{Accepted 2015 February 17.  Received 2015 February 16; in original form 2015 January 8}

\begin{document} 

\maketitle 

\begin{abstract} 
Ground-based photometric observations from WASP have been calibrated, scaled and combined with \Kepler observations of the \dsct star KIC~7106205, allowing us to extend the time base of the study of the unexplained amplitude and frequency variation of a single pressure mode at $\nu=13.3942$~d$^{-1}$ by 2~yr. Analysis of the combined data sets, spanning  6~yr, show that the amplitude modulation in KIC~7106205 has a much larger range than a previous study of the {\it Kepler} data alone indicated. The single pressure mode decreased from $11.70 \pm 0.05$~mmag in 2007, to $5.87 \pm 0.03$~mmag in 2009, and to $0.58 \pm 0.06$~mmag in 2013. Observations of the decrease in mode amplitude have now been extended back 2~yr before the launch of {\it Kepler}. With observations over a longer time span, we have been able to further investigate the decrease in mode amplitude in KIC~7106205 to address the question of mode amplitude stability in $\delta$~Sct stars. This study highlights the usefulness of the WASP data set for extending studies of some {\it Kepler} variable stars.
\end{abstract} 



\section{Introduction}

The {\it Kepler} mission photometric data set spans 4~yr for over 190\,000 stars, with precision in the measurement of the amplitudes of pulsating stars of the order of a few $\umu$mag. While these data provide unprecedented views of the light variations in variable stars, many stars are not purely periodic, and we wonder what happened for these stars before the {\it Kepler} data set started and after the {\it Kepler} data set ended. In this paper we demonstrate how including ground-based photometric data, specifically from the Wide Angle Search for Planets (WASP) project, alongside \Kepler data allows us to study the multi-periodic \dsct star KIC~7106205 (TYC~$3129$-$879$-$1$; $1$SWASPJ$191157.46$+$424022.6$) over a longer time span than the 4~yr of the {\it Kepler} data set -- a total of 6~yr in this case. Our aim is to investigate the unusual frequency and amplitude modulation in only a single pulsation frequency, discovered by \cite{Bowman2014}, and to test if this behaviour is present in other observations of this star, specifically the WASP data set.

Periodic changes in the surface brightness and measurements of radial velocities of pulsating stars allow us to probe stellar interiors and has driven remarkable advances in understanding the mechanisms that drive pulsations in variable stars, especially \dsct stars (e.g. \citealt{Dupret2004, Dupret2005}). The \dsct stars lie at the intersection of the main-sequence and the classical instability strip on the Hertzsprung--Russell diagram \citep{Uytterhoeven2011}. The pulsations are driven by the $\kappa$-mechanism, in which changes in opacity set up a piston-like change in the radius of a shell of gas in the He~{\sc{ii}} ionisation zone \citep{Chevalier1971a}. Consequently, low-order p~modes are excited, with typical periods of 15~min to 5~hr \citep{Uytterhoeven2011}. See \cite{Breger2000a} and \cite{Murphy_PhD} for reviews of \dsct stars.

Pulsational frequency and/or amplitude modulation are observed in a variety of stellar types, e.g. solar-like stars \citep{Chaplin2000, Chaplin2007}, \dsct stars \citep{Breger2014, Bowman2014}, roAp stars \citep{Kurtz1997, Holdsworth2014b}, and white dwarfs \citep{Winget1991}. For stochastically driven pulsators, the short mode lifetimes are easy to explain, given the nature of the driving and damping mechanisms that are competing within the star. However, no theory exists that can explain the diversity of behaviours observed, especially in the classical pulsators, e.g. the \dsct stars (see \cite{Breger2014} and references therein).

\subsection{The \Kepler mission}

The \Kepler space telescope was launched in 2009 March and observed more than 190\,000 stars at high photometric precision and high duty-cycle \citep{Koch2010}. The primary goal of the mission was to observe Earth-like planets within the habitable zone of their host star using the transit method \citep{Borucki2010}. A total of 4~yr (1470~d) of observations were collected, covering a 115~deg$^2$ field of view in the constellations of Cygnus and Lyra. Observations, taken in the passband of $420-900$~nm \citep{Koch2010}, were made in two modes, long cadence (LC) of 29.5~min, and short cadence (SC) of 58.5~s \citep{Gilliland2010}. These excellent data have also proved invaluable for studying pulsating stars. A frequency resolution of 7.9~nHz (0.00068~d$^{-1}$) is obtained when calculating the Fourier transform of all 1470~d of \Kepler observations, so only since the end of the main mission has it been possible to study frequency modulation in \dsct stars at this precision, and to study amplitude modulation over this time span. 

\subsection{The WASP project}

The WASP project is a two-site wide-field survey for transiting exoplanets \citep{Pollacco2006}. The instruments are located at the Observatorio del Roque de los Muchachos on La Palma and at the Sutherland Station of the SAAO, and achieved first light in 2003 and 2005, respectively. The instruments consist of eight 200-mm, f/1.8 Canon telephoto lenses mounted in a $2\times4$ configuration. Each is backed by an Andor CCD of $2048\times2048$ pixels, allowing a pixel size of about 14~arcsec. Observations are made through broad-band filters of $400-700$~nm for the eight lenses. The data pass through a reduction pipeline correcting for primary and secondary extinction, the colour-response of the instrument, the zero point and atmospheric extinction. The pipeline is optimised for G stars. The data are also corrected for instrumental systematics using the {\sc{sysrem}} algorithm of \cite{Tamuz2005}.

The observing strategy of WASP provides two consecutive 30~s exposures at a given pointing, before moving to the next available field; fields are typically revisited every 10~min. Such a strategy has enabled the discovery of many types of variable stars, from low-frequency binary stars \citep{Smalley2014}, to high-frequency pulsating A stars \citep{Holdsworth2014a}. For further details of the WASP project and the techniques used for the detection of pulsations, we refer the reader to \cite{Pollacco2006} and \cite{Holdsworth2014a}, respectively.

\subsection{Mode-coupling in \dsct stars}

The \dsct stars demonstrate diverse pulsational behaviour, but no theory exists to describe all the phenomena observed, especially variable pulsation amplitudes. Mode-coupling is predicted between frequencies in \dsct stars \citep{Dziembowski1982}, specifically parametric resonance in which the instability of a linearly-driven mode at $\nu_1$ causes the growth of two modes at $\nu_2$ and $\nu_3$, such that $\nu_1 \approx \nu_2 + \nu_3$. The most likely outcome of these nonlinear effects is the decay in amplitude of a linearly-driven p~mode, as such modes usually have low radial orders, and the growth of two g~modes \citep{Dziembowski1982, Nowakowski2005}. These g~modes can become trapped in a pulsation cavity close to the core of the star and are therefore invisible at the stellar surface. Thus, these are considered {\it internal} g~modes. 

Observationally, the resonant mode-coupling predictions of \cite{Dziembowski1982} and \cite{Nowakowski2005} are difficult to test using broad-band photometry alone, as high-degree (high-$\ell$) p~modes and internal g~modes have small amplitudes at the surface of the star. On the other hand, studies of variable amplitudes of modes of low-degree ($\ell \leq 2$) are relatively easy as a consequence of their higher visibility.


\section{Previous study of KIC~7106205}

KIC~7106205 is a multi-periodic \dsct star that was studied by \cite{Bowman2014} using \Kepler data. They tracked amplitude and phase at fixed frequency for 16 significant pulsation modes over the 1470~d time span of the {\it Kepler} data set. All pulsation frequencies were found to be remarkably stable except for $\nu_{\rm mod} =13.3942$~d$^{-1}$, which decreased in amplitude from $5.87 \pm 0.03$~mmag in 2009 to $0.58 \pm 0.06$~mmag in 2013, corrected for the \Kepler integration time (see Eqn.~\ref{equation: integration correction}). The pulsation constant of $\nu_{\rm mod}$ indicates it is likely a third or fourth radial overtone mode. Higher frequencies (i.e. higher overtones) were observed to be stable, thus the observed modulation was not constrained to the surface of the star \citep{Bowman2014}.

\begin{table}
	\centering
	\caption[]{Stellar parameters listed for KIC~7106205 in the \textit{Kepler} Input Catalogue (KIC; \citealt{Brown2011}) and the revised values given in \citet{Huber2014}.}
		\begin{tabular}{c c c c c c c}
		\hline
		\multicolumn{1}{c}{ } & \multicolumn{1}{c}{T$_{\mathrm{eff}}$} & \multicolumn{1}{c}{$\log g$ } & \multicolumn{1}{c}{[Fe/H]} & \multicolumn{1}{c}{K$_{\mathrm{p}}$ mag} \\
		\multicolumn{1}{c}{ } & \multicolumn{1}{c}{(K)} & \multicolumn{1}{c}{(cm~s$^{-2}$)} & \multicolumn{1}{c}{(dex)} & \multicolumn{1}{c}{(mag)} \\
		\hline
		KIC & 6960 $\pm$ 150 & 4.05 $\pm$ 0.15 & -0.01 $\pm$ 0.15  & 11.46  \\ 
		revised & 6900 $\pm$ 140 & 3.70 $\pm$ 0.15 & 0.32 $\pm$ 0.15  &   \\ 
		\hline
		\end{tabular}
	\label{table: stellar parameters}
\end{table}

The loss of mode energy of $\nu_{\rm mod}$ was not observed to excite any new frequencies or to be transferred to any existing frequencies. Therefore, it was concluded that energy was lost to a damping region within the star or transferred to either high-$\ell$ p~modes or two internal g~modes via the parametric resonance instability. Both outcomes would result in modes that are invisible at the stellar surface using broad-band photometry.

\begin{figure*}
	\centering
	\includegraphics[width=0.85\textwidth]{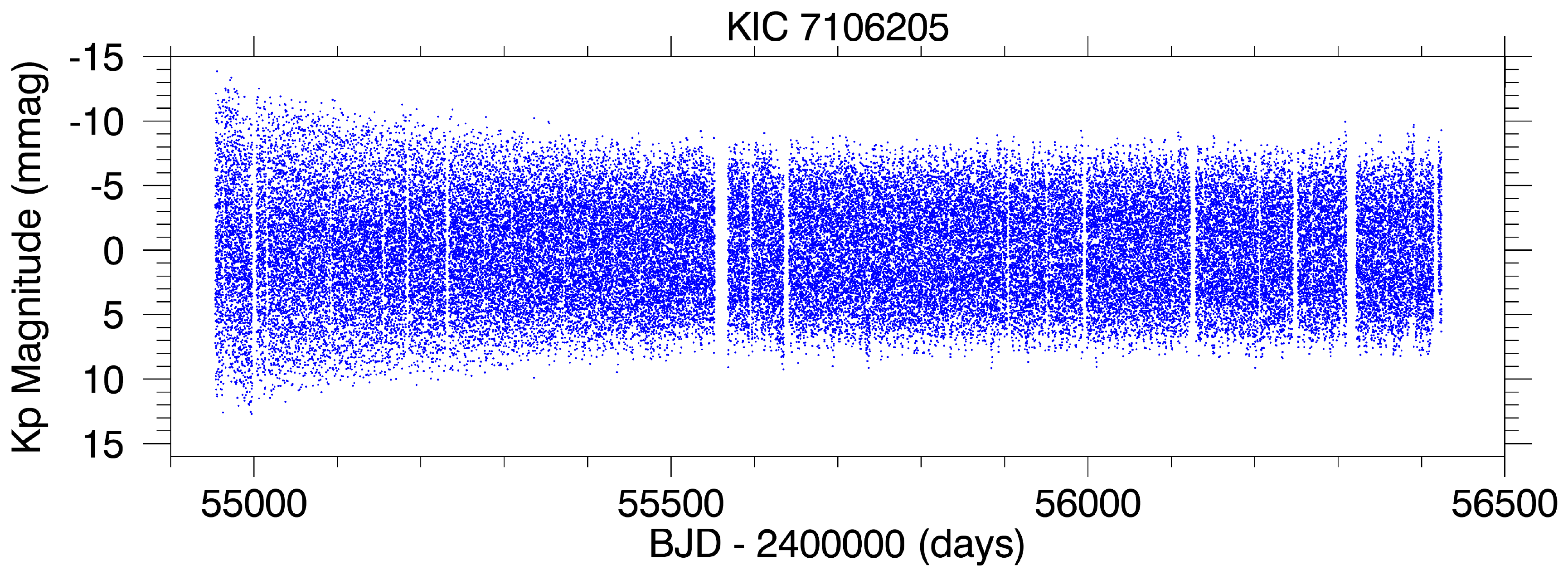}
	\includegraphics[width=0.85\textwidth]{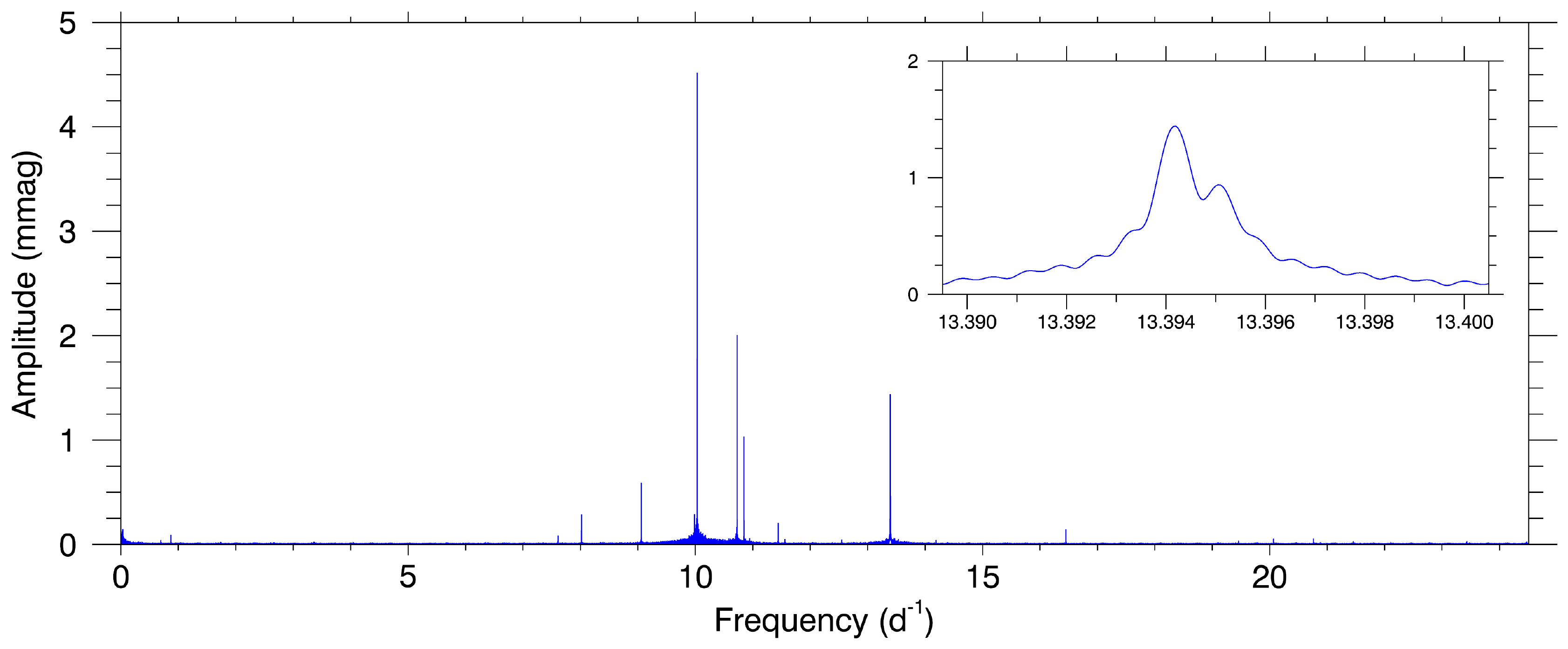}
	\caption{The 1470~d \Kepler light curve for the \dsct star KIC~7106205 is given in the top panel and the bottom panel is the amplitude spectrum, calculated out to the LC Nyquist frequency. The sub-plot in the bottom panel shows a zoomed-in view of the modulated mode $\nu_{\rm mod} = 13.3942$~d$^{-1}$.}
	\label{figure: 7106205}
\end{figure*}

The 1470-d light curve for KIC~7106205 and the subsequent amplitude spectrum are given in the top and bottom panels of Fig.~\ref{figure: 7106205}, respectively. The amplitude spectrum illustrates that KIC~7106205 contains only a modest number of pulsation frequencies. Table \ref{table: stellar parameters} provides the stellar parameters of KIC~7106205 from both the \Kepler Input Catalogue (KIC; \citealt{Brown2011}) and the revised values given in \citet{Huber2014}.


\section{Method: combining the data}

The instrumental differences between \Kepler and WASP require the data to be corrected before a direct comparison of the data sets can be made. The pulsation amplitude correction from the difference in integration times for each data set is calculated using 

\begin{equation}
	A = A_{0}~{\rm sinc}\left(\frac{\pi}{n}\right) ,
	\label{equation: integration correction}
\end{equation}

\noindent where $A$ and $A_{0}$ are the observed and corrected amplitudes, respectively, and $n$ is the number of data points per pulsation cycle \citep{Murphy_PhD}.

\subsection{Using the HADS star KIC~9408694 for calibration}

\begin{figure*}
	\centering
	\includegraphics[width=0.8\textwidth]{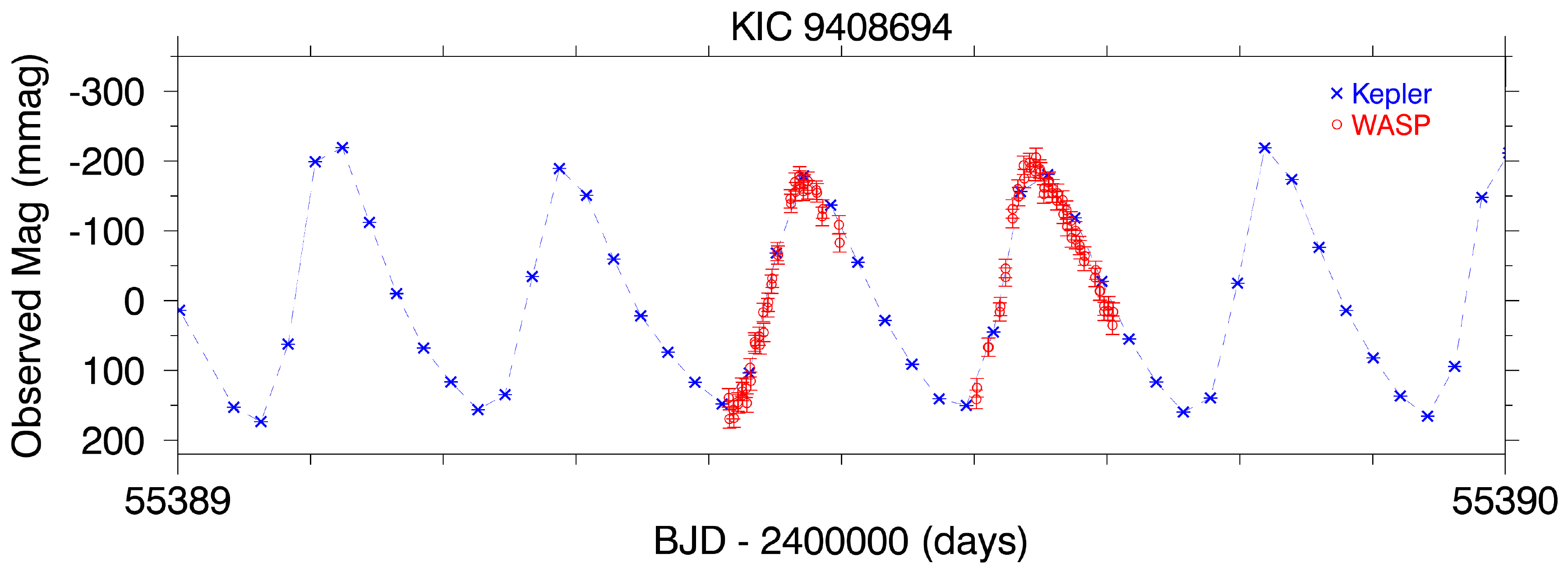}
	\includegraphics[width=0.48\textwidth]{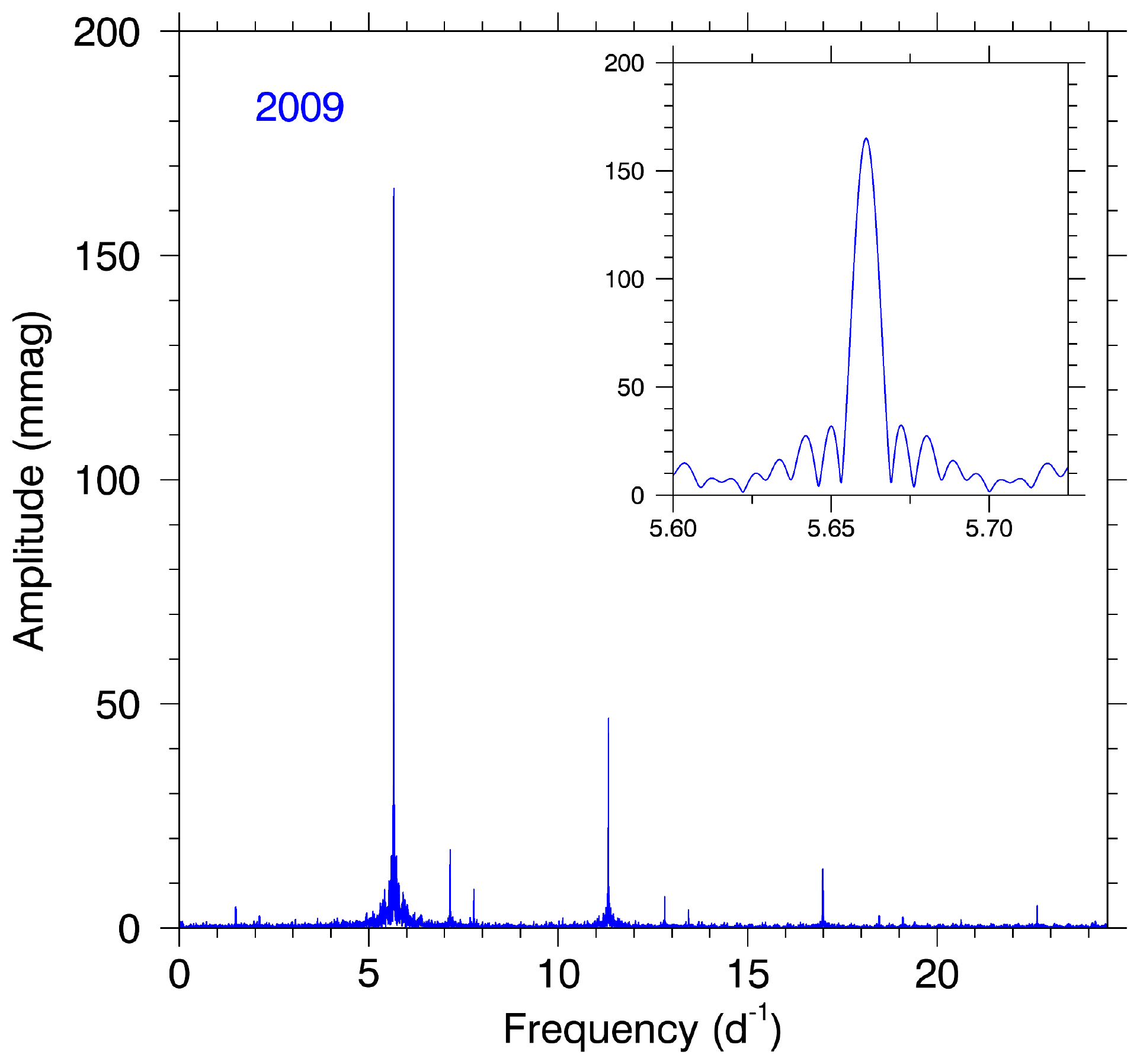}
	\includegraphics[width=0.48\textwidth]{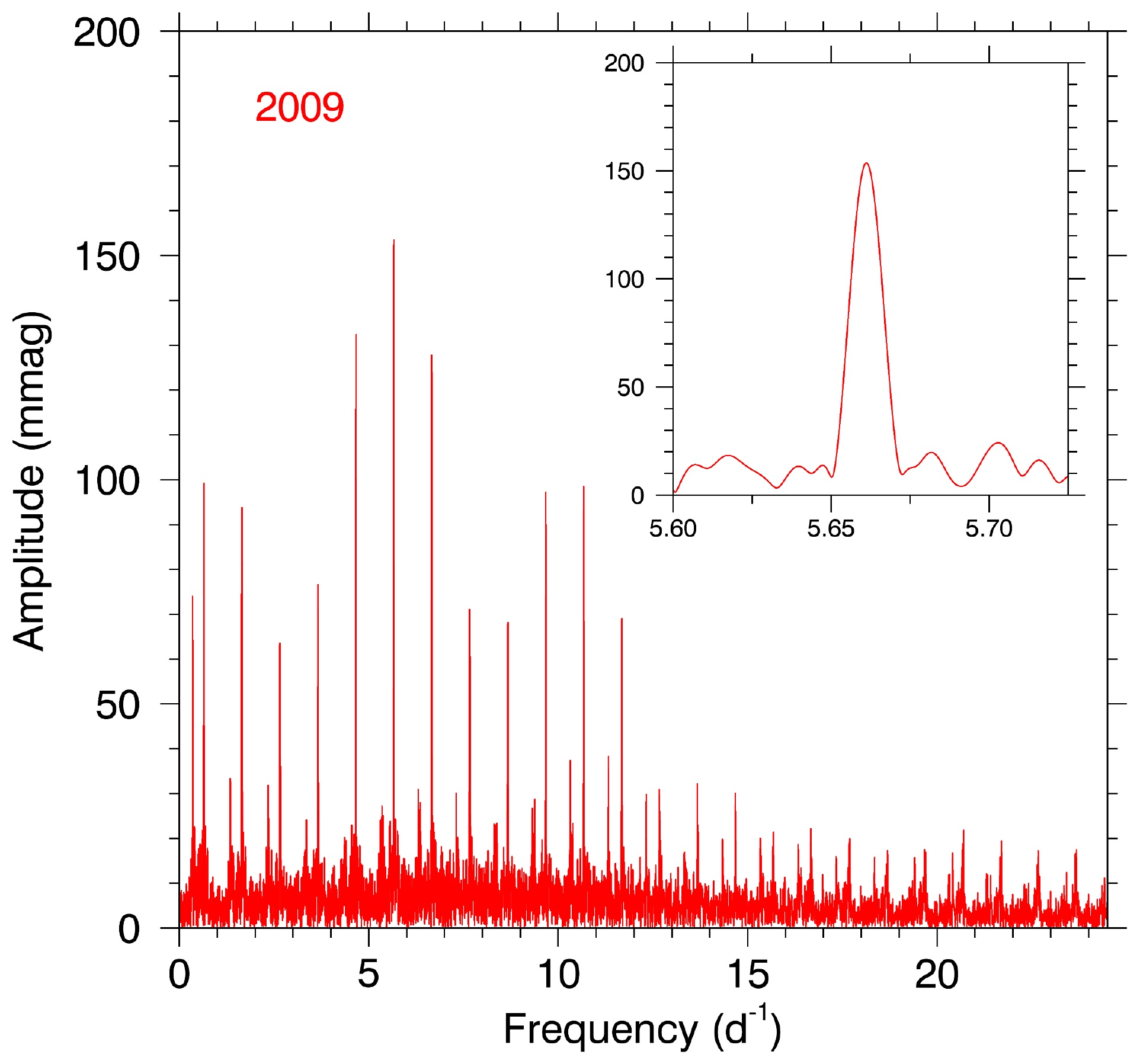}
	\includegraphics[width=0.48\textwidth]{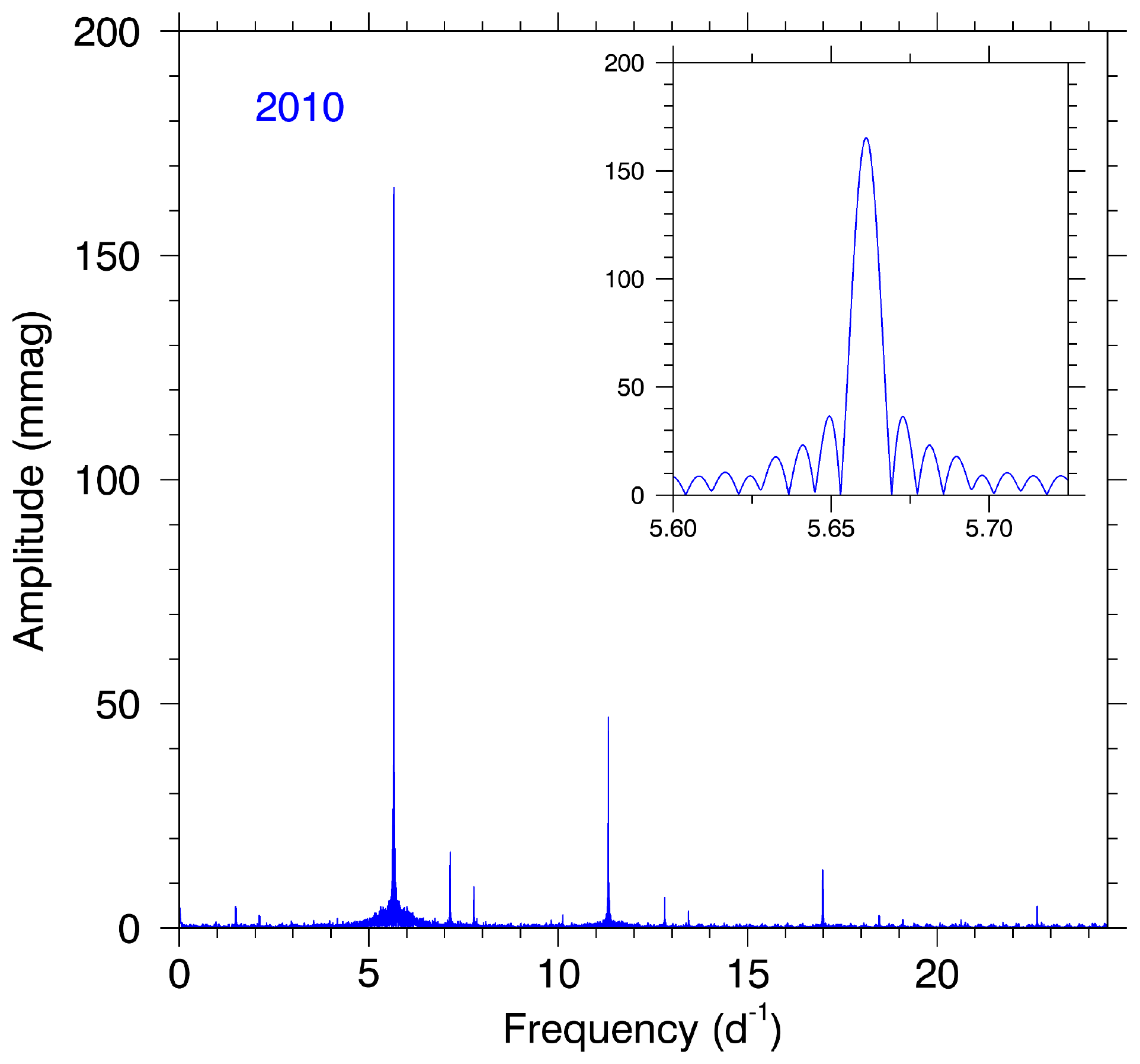}
	\includegraphics[width=0.48\textwidth]{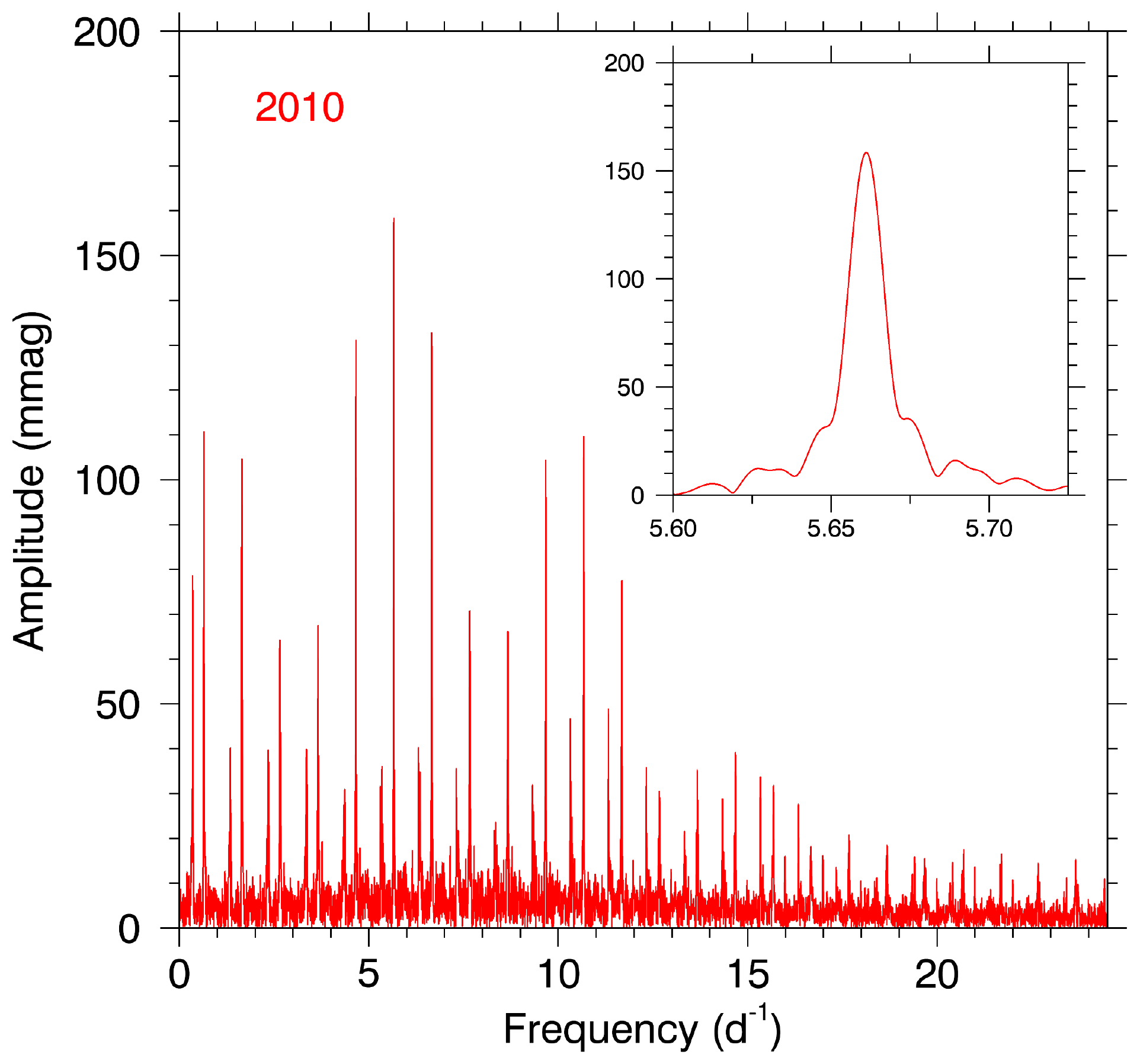}
	\caption{A 1-d sample showing simultaneous WASP and \Kepler observations of the HADS star KIC~9408694 is given in the top panel. The left and right columns are the amplitude spectra, calculated out to the \Kepler LC Nyquist frequency, for the \Kepler (blue) and WASP (red) time series, respectively, in which 2009 and 2010 are the top and bottom rows. The sub-plots in the bottom panels show a zoomed-in view of the mode at $\nu = 5.6611$~d$^{-1}$, used for calculating the \Kepler and WASP passband differences.} 
	\label{figure: HADS}
\end{figure*}

A calculation of the effect on the measurement of pulsation amplitude caused by the passband differences between {\it Kepler} and WASP was conducted using a high-amplitude \dsct (HADS) star. The HADS stars are a subgroup of \dsct stars that are found in the central region of the instability strip \citep{McNamara2000}. They generally have pulsation amplitudes greater than approximately 0.3~mag and pulsate in fundamental and first overtone radial modes (e.g. see \citealt{Balona2012a}). A HADS star was chosen for the passband calibration, as the signal-to-noise ratio (SNR) is extremely high for high-amplitude pulsations. Moreover, the particular pulsation mode did not vary in amplitude over time span of the 4~yr {\it Kepler} data set. The HADS star used was KIC~9408694, which has a well-resolved dominant peak at $5.6611$~d$^{-1}$. A 1-d sample of \Kepler and WASP over plotted light curves and the amplitude spectra of KIC~9408694 in 2009 and 2010 for both instruments are shown in Fig.~\ref{figure: HADS}. 

The HADS star is not isolated in the WASP aperture, but the background objects are 3 magnitudes fainter and added only a small amount of flux to the photometry. The dilution suffered by the target star can be calculated by comparing the flux of the target star and contaminating stars such that
\begin{equation}
	Dilution = \left[ 1-\left(\frac{F_{\rm T}}{F_{\rm T}+F_{\rm C}}\right)\right]\times100~,
	\label{equation: dilution correction}
\end{equation}
where $F_{\rm T}$ is the flux of the target and $F_{\rm C}$ is the combined flux of the contaminating stars (Holdsworth Ph.D. Thesis, {\it in prep}). The dilution for the HADS star KIC~9408694 was calculated to be 14~per~cent (i.e. the observed amplitude is reduced by 1.1400). 

\begin{figure*}
	\centering
	\includegraphics[width=0.8\textwidth]{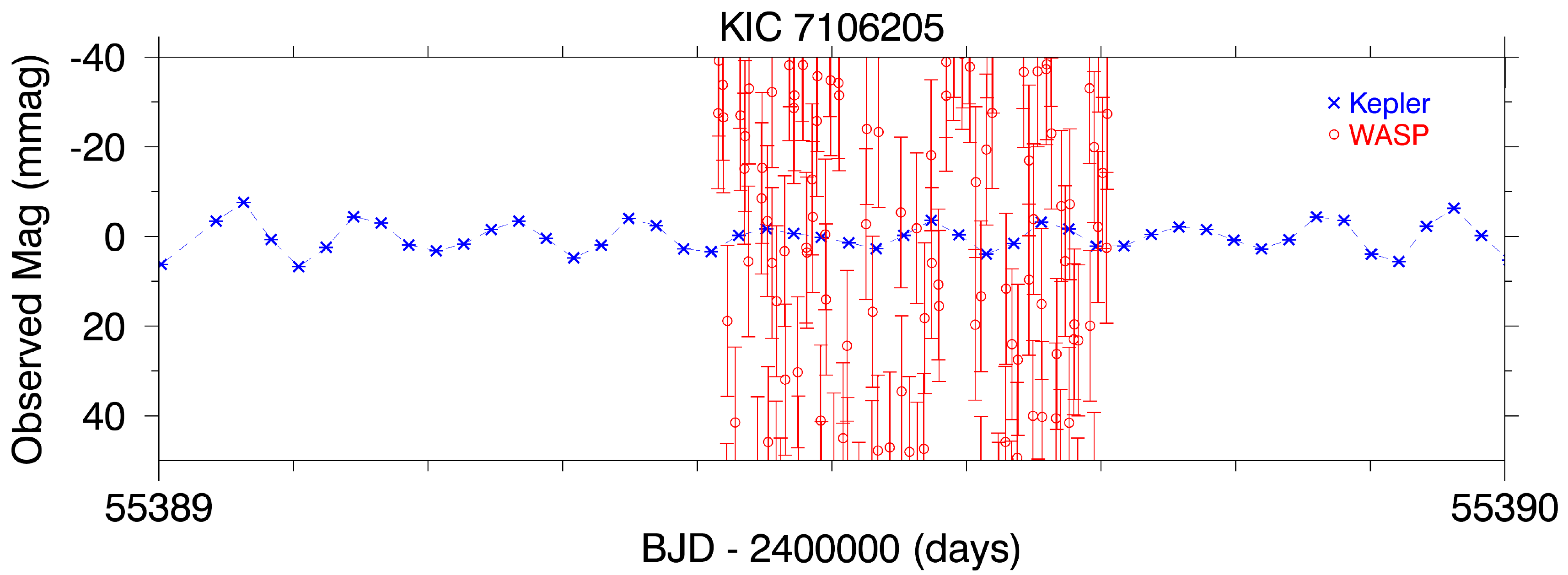}
	\includegraphics[width=0.48\textwidth]{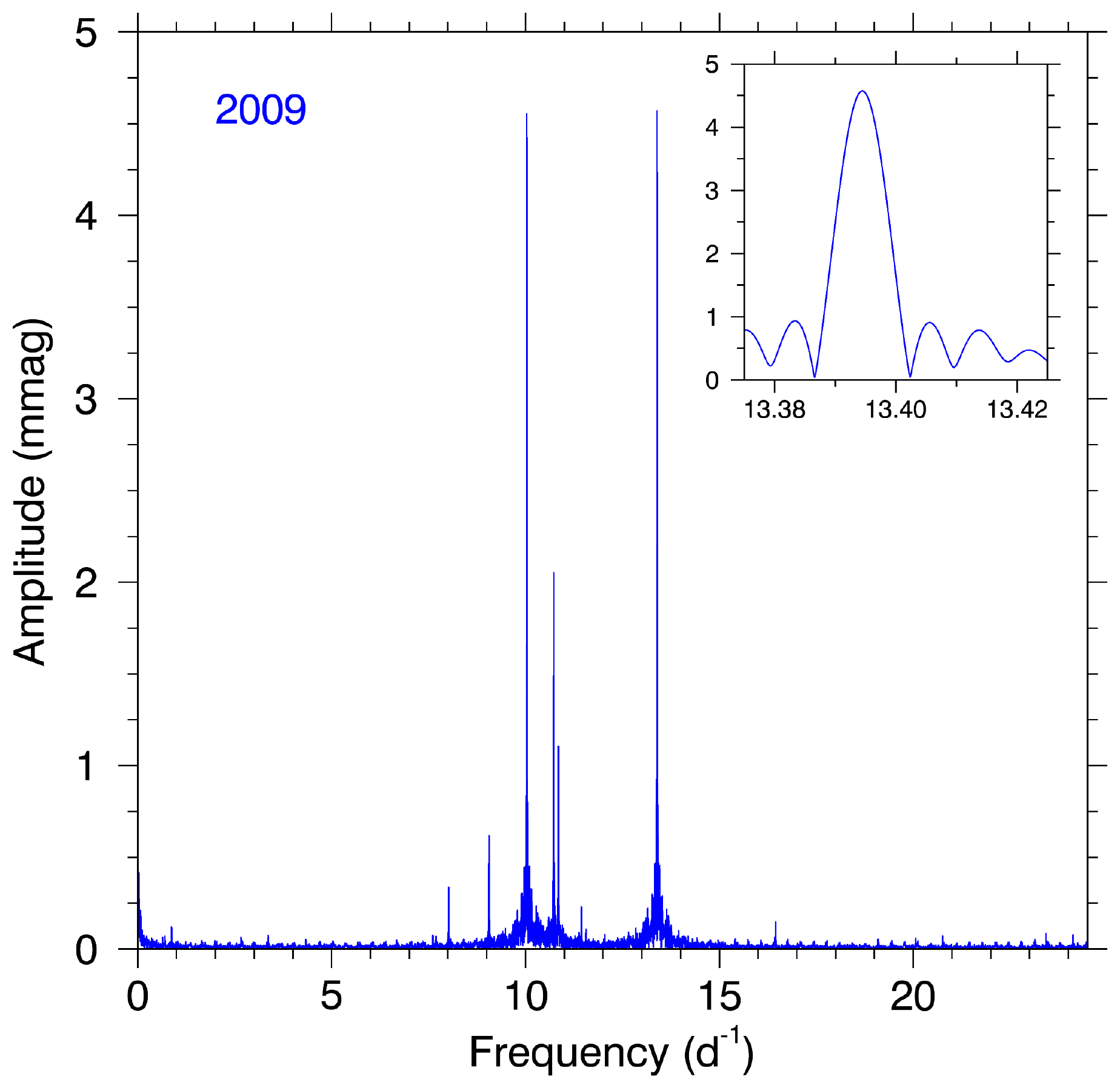}
	\includegraphics[width=0.48\textwidth]{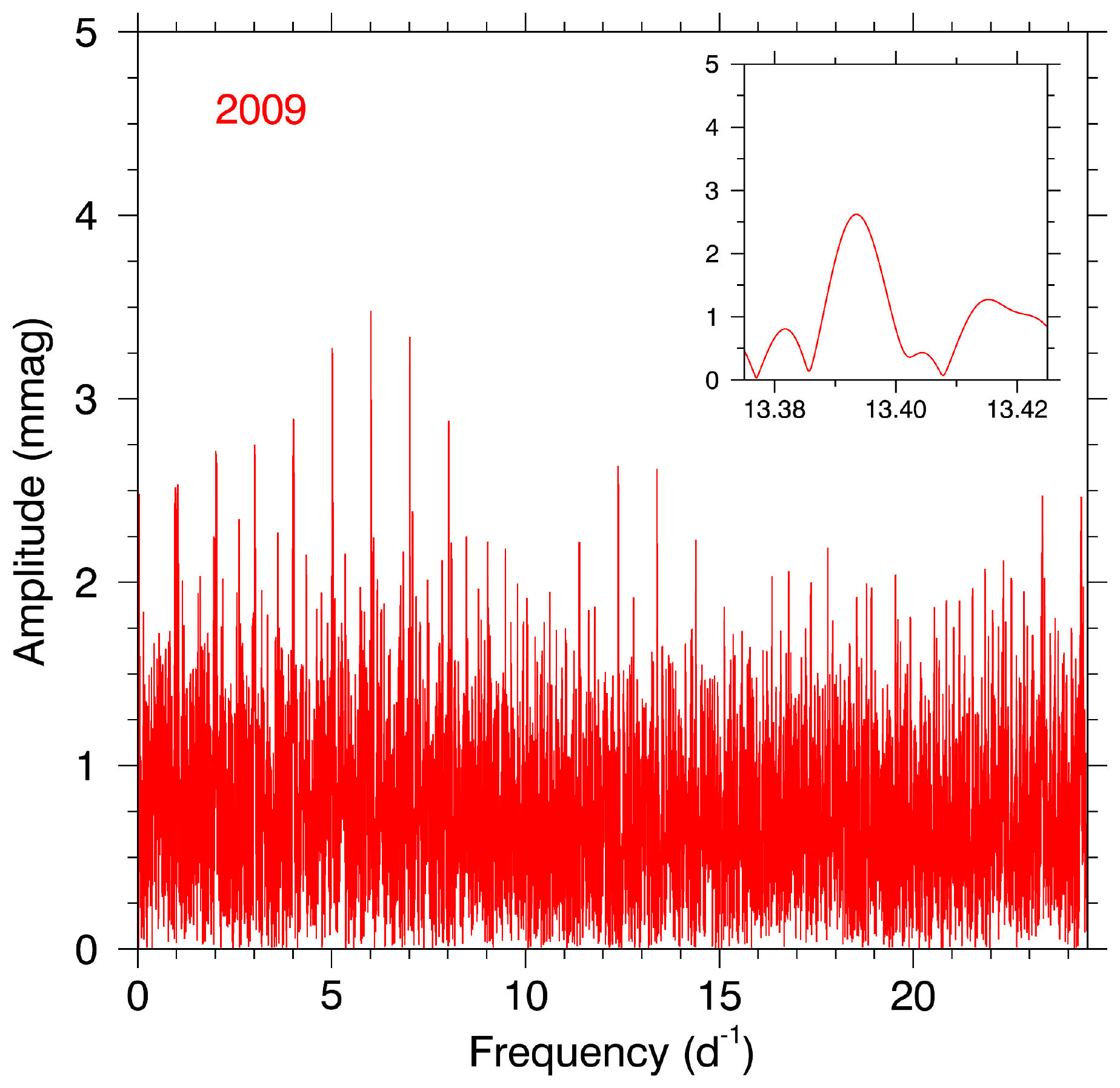}
	\includegraphics[width=0.48\textwidth]{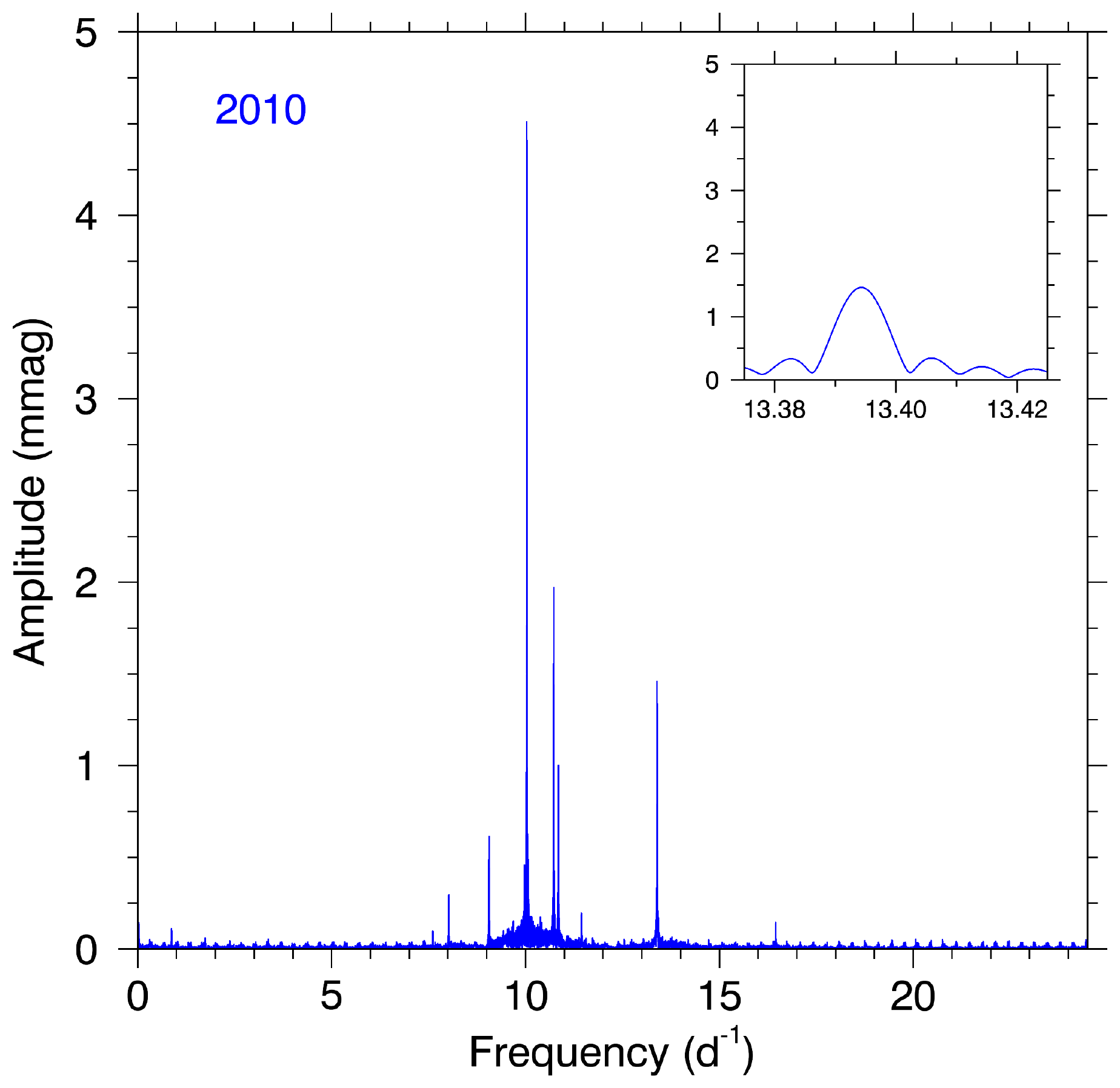}
	\includegraphics[width=0.48\textwidth]{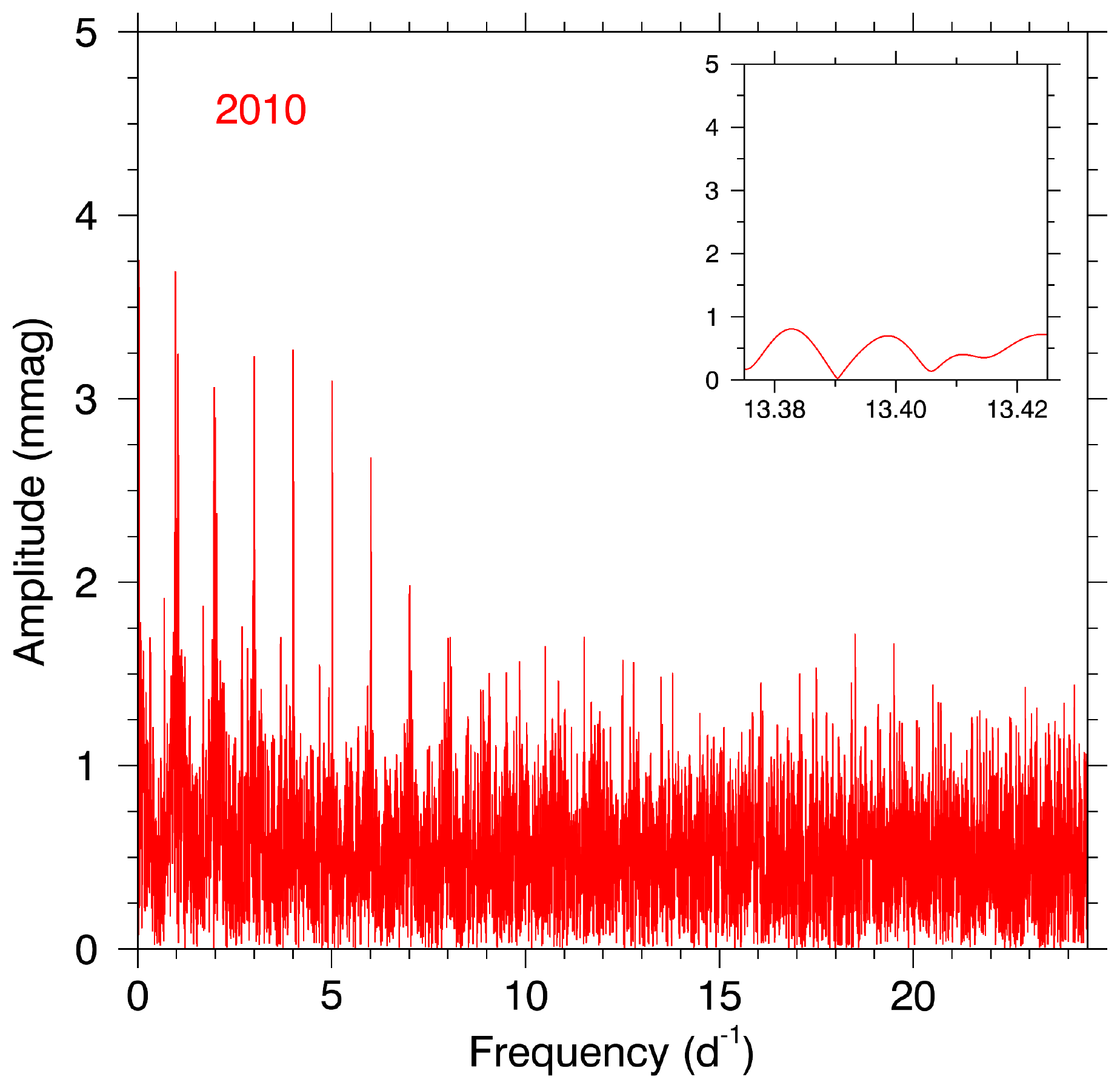}
		\caption{A 1-d sample showing simultaneous WASP and \Kepler observations of the \dsct star KIC~7106205 is given in the top panel, in which the larger scatter of WASP observations due to a higher point noise is clearly seen. The left and right columns are the amplitude spectra, calculated out to the \Kepler LC Nyquist frequency, for the \Kepler (blue) and WASP (red) time series, respectively, in which 2009 and 2010 are the top and bottom rows. The sub-plots in panels show a zoomed-in view of the modulated mode $\nu_{\rm mod} = 13.3942$~d$^{-1}$.}
	\label{figure: DSCT}
\end{figure*}

Simultaneous WASP and \Kepler observations of KIC~9408694 were made in 2009 and 2010. The data from WASP and \Kepler were truncated to the same time period and analysed to determine their pulsation amplitudes, $A_{0_{\rm W}}$ and $A_{0_{\rm K}}$, respectively. The difference in the derived amplitudes, (after optimising using linear least-squares, correcting for dilution in the WASP data and integration time effects in both data sets), is solely a result of the difference in the passbands and the ratio allows one data set to be scaled for comparison to the other. The average of the ratio $(A_{0_{\rm K}} / A_{0_{\rm W}})$ for the two years was $0.9242$, i.e. the \Kepler data show amplitudes 7.58~per cent smaller than those of the WASP data due to the filter differences, which is an expected result for an A star given the filter responses.

\subsection{Application to the \dsct star KIC~7106205}

\begin{figure}
	\centering
	\includegraphics[width=0.75\linewidth]{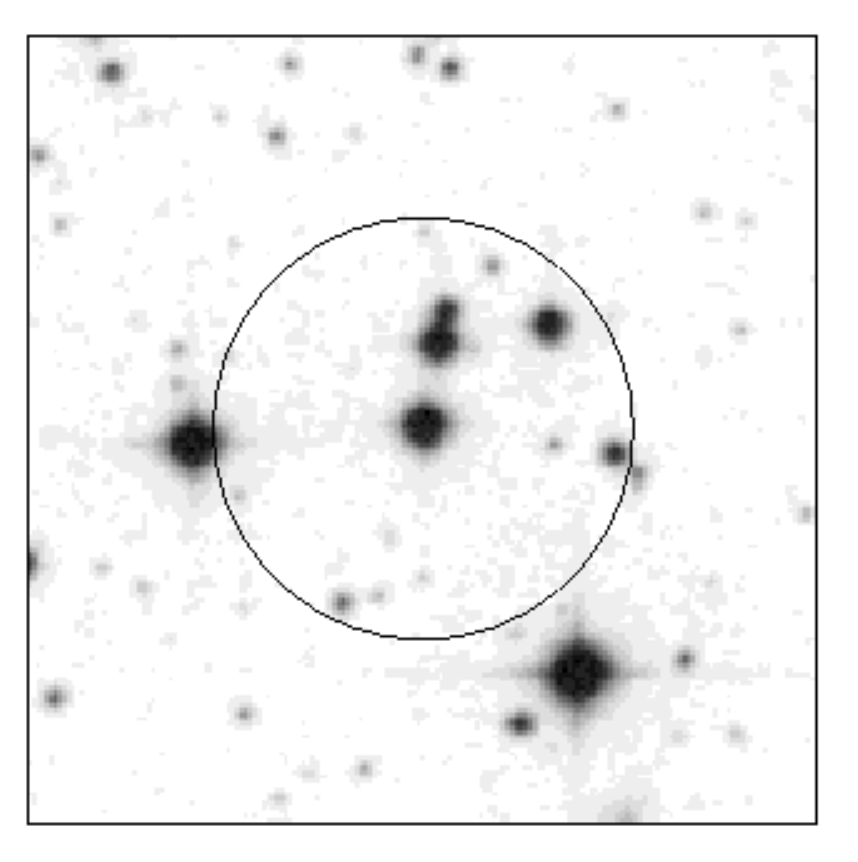}
	\caption{The photometric aperture (dark circle) for KIC~7106205 in the WASP data showing multiple sources of contamination. The aperture is 48\arcsec~ (3.5 WASP pixels) in radius. Image from DSS.}
	\label{figure: WASP DSCT field}
\end{figure}

In the WASP data, KIC~7106205 was observed for three seasons, 2007, 2009 and 2010. Due to the large pixel size of the WASP instrument, the star suffers from contamination from other sources in the photometric aperture (see Figs \ref{figure: DSCT} and \ref{figure: WASP DSCT field}). Therefore, a dilution correction factor must also be calculated before comparison can be made with \Kepler data. Simultaneous observations by WASP and \Kepler of KIC~7106205 in 2009 show the $\nu_{\rm mod}$ peak.

To perform the comparison, the \Kepler data were truncated to the same time period as the WASP data, and the pulsation peak $\nu_{\rm mod}$ was extracted from both data sets, optimised by linear least-squares and then corrected for the different integration times as described previously. This gave a corrected WASP amplitude of $A_{0_{\rm W}}=2.77$~mmag and a \Kepler amplitude of $A_{0_{\rm K}}=5.18$~mmag. 

To calculate the dilution effect in KIC~7106205, the \Kepler peak was transformed into the WASP passband, giving the expected WASP amplitude if no dilution occurred, $A_{\rm exp}$. The ratio, therefore, between $A_{\rm exp}$ and $A_{0_{\rm W}}$  is a result of the dilution of other stars in the aperture; this is calculated to be $2.0212$. Alternatively, if Eqn.~\ref{equation: dilution correction} is used, a dilution factor of 1.7098 is obtained. The difference in this value and the direct comparison of the two data sets is possibly due to the bleeding of light from other nearby bright stars into the extracted pixels. Therefore, we conclude that the most reliable method is to compare the pulsation amplitudes from both instruments, as this includes any residual effect and corrects for it along with the dilution.

Fig. \ref{figure: DSCT} shows the 2009 and 2010 \Kepler and WASP data used for the calibration study of KIC~7106205. The top panel shows an example of simultaneous observations by WASP and \Kepler in 2009. The bottom-left column shows the amplitude spectra of the \Kepler 2009 and 2010 data, with the bottom-right column showing the amplitude spectra of the same two of the three seasons of available WASP data, as in the \Kepler data.

Unfortunately, the 2010 WASP data do not show the $\nu_{\rm mod}$ peak. The amplitude of $\nu_{\rm mod}$ in the 2010 \Kepler data is $1.46$~mmag, and correcting this for integration time gives $A_{0_{K}} = 1.66$~mmag. The passband difference and dilution factor in KIC~7106205 of $2.0212$, and the WASP integration time correction means that we would expect an amplitude of $0.87$~mmag in the 2010 WASP data. Despite being above the nominal detection limit of $0.50$~mmag \citep{Holdsworth2014a}, $\nu_{\rm mod}$ is not observed in the 2010 WASP data due to higher photometric noise.

\begin{figure}
	\centering
	\includegraphics[width=0.99\linewidth]{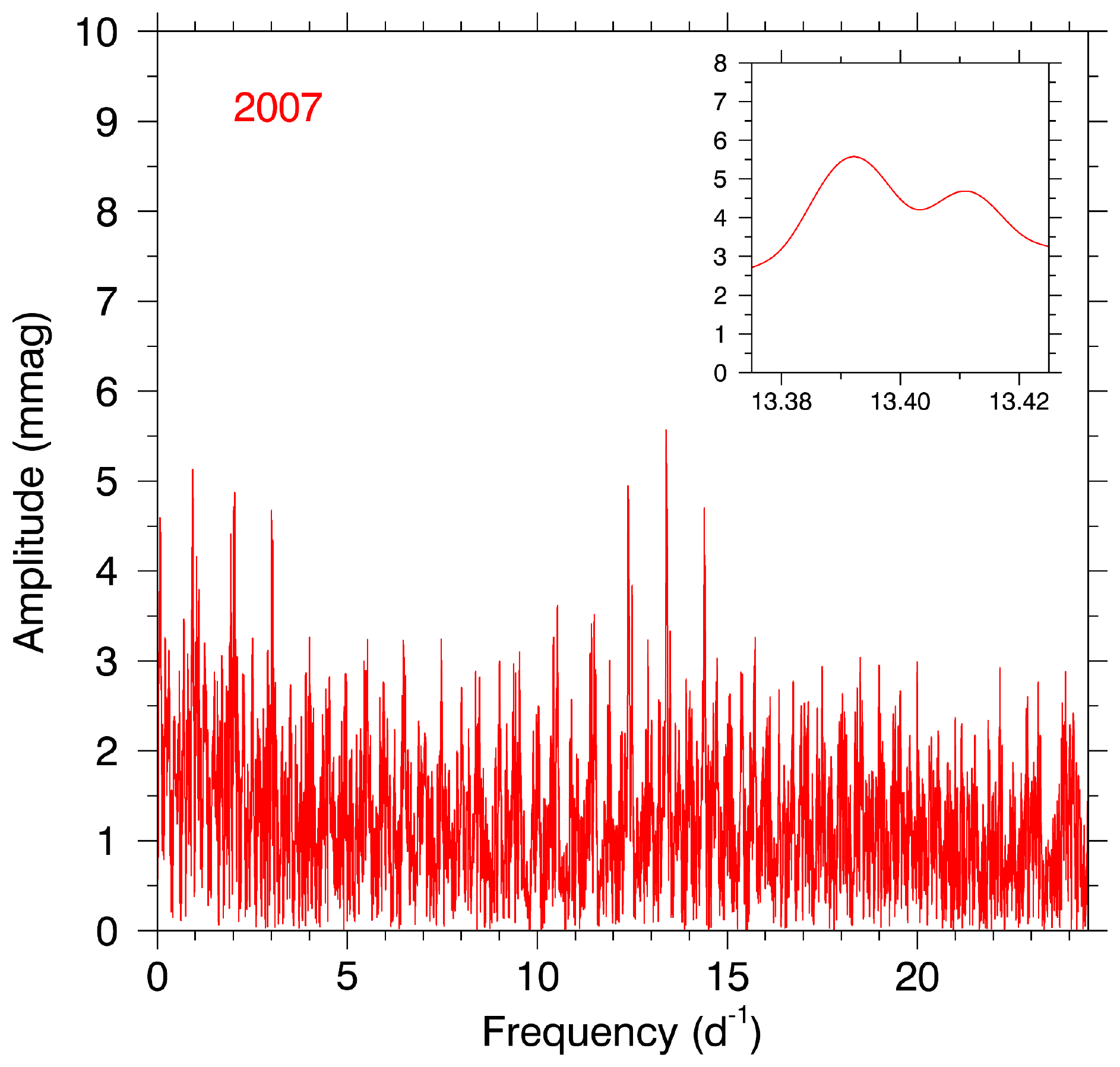}
	\caption{The amplitude spectrum of the 2007 WASP data for KIC~7106205, calculated out to the \Kepler LC Nyquist frequency. The sub-plot shows a zoomed-in view of $\nu_{\rm mod}$. Note the difference in y-axis scale compared to the amplitude spectra given in Fig.~\ref{figure: DSCT}, illustrating the large decrease in mode amplitude between 2007 and 2009.}
	\label{figure: WASP DSCT 2007 FT}
\end{figure}

The 2007 WASP data set for KIC~7106205 is 66.1~d in length and consists of 3322 data points. Fig.~\ref{figure: WASP DSCT 2007 FT} shows the amplitude spectrum of the 2007 WASP data, from which the amplitude of $\nu_{\rm mod}$ was extracted, optimised by linear least-squares at a fixed frequency of $13.3942$~d$^{-1}$ and then corrected using the methodology described previously. Due to the higher levels of noise present in WASP data compared to \Kepler data, it is not possible to track amplitude changes in all pulsation modes and so we were unable to perform an analysis of all the frequencies listed in \cite{Bowman2014}. However, $\nu_{\rm mod}$ has a sufficiently high amplitude that it can be extracted from the 2007 WASP observations and calibrated to find its corrected amplitude. 

\section{Results}

\begin{figure*}
	\centering
	\includegraphics[width=0.75\linewidth,angle=0]{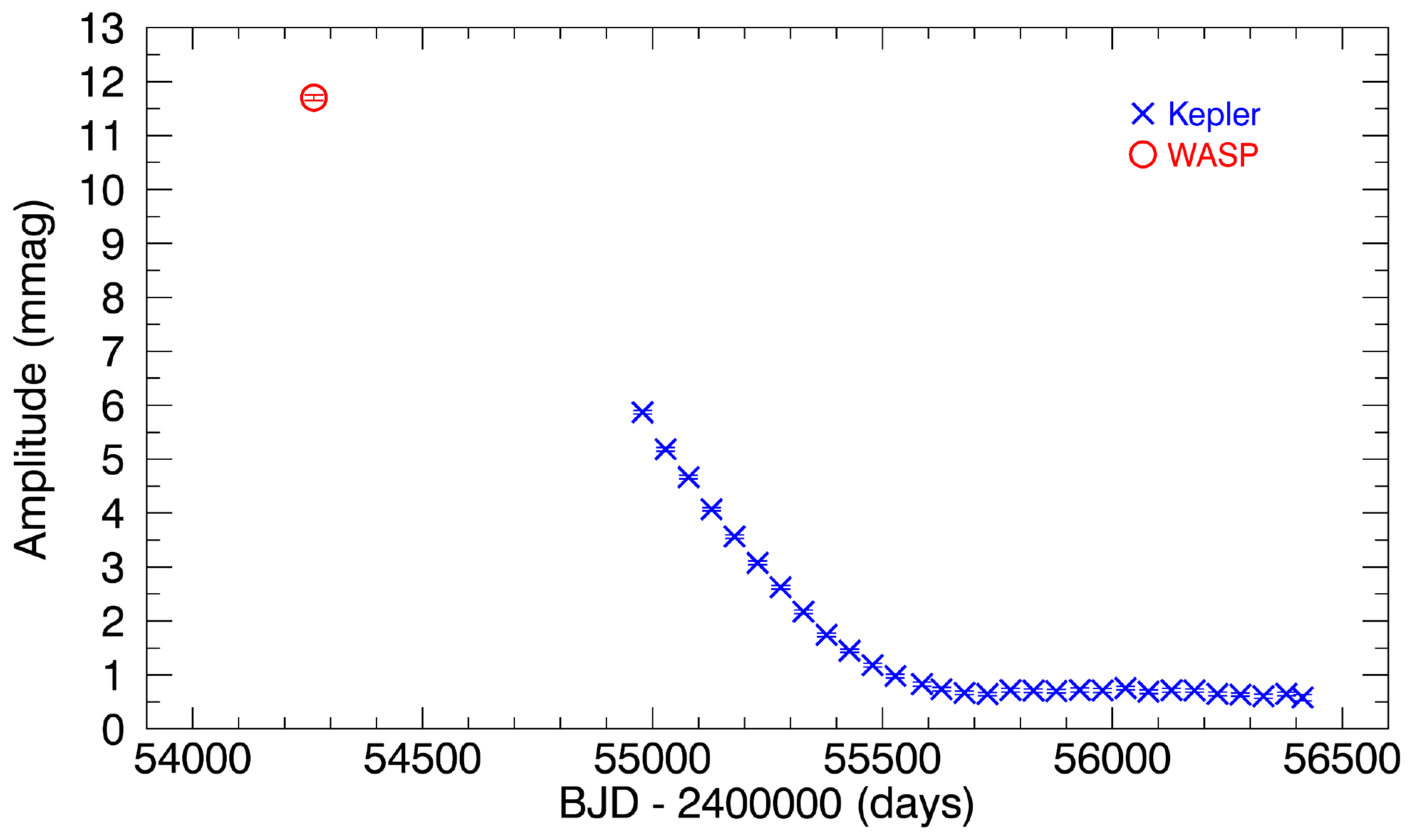}	
	\caption{Corrected and calibrated values of amplitude of $\nu_{\rm mod}$ in KIC~7106205 for \Kepler (in 50~d bins) and WASP (single 66.1-d bin) data, marked as blue crosses ($\times$) and a red circle ($\circ$), respectively. 1$\sigma$ errors are calculated from the least-squares fit at fixed frequency of $13.3942$~d$^{-1}$, but are generally smaller than the data points.}
	\label{figure: max plots}
\end{figure*}

The 66.1~d of 2007 WASP data for KIC~7106205 were combined into a single data bin, so that the highest possible frequency resolution ($0.015$~d$^{-1}$) was obtained when calculating a discrete Fourier transform. The amplitude was obtained by optimising the amplitude and phase using a least-squares fit, which was normalised to the centre of the 1470~d \Kepler data set, specifically $t_0 = 2\,455\,688.77$ (BJD), and at the fixed frequency of $13.3942$~d$^{-1}$. This ensures that values of amplitude can be compared to the analysis of \cite{Bowman2014}.
	
The resultant amplitude from the least-squares fit was calibrated, as described in previous sections for integration time, dilution from other stars and passband differences (in that order).  A corrected amplitude of $11.70 \pm 0.05$~mmag was obtained for $\nu_{\rm mod}$ and is shown graphically in Fig.~\ref{figure: max plots}, which contains the single calibrated WASP data point alongside 1470~d of \Kepler data for KIC~7106205 adapted from \cite{Bowman2014}. The inclusion of the WASP data point illustrates that the amplitude of $\nu_{\rm mod}$ appears to have been steadily decreasing since at least 2007.

It should also be noted that if a nonlinear least-squares fit of the 2007 WASP data of KIC~7106205 is performed, values of $\nu = 13.3924$~d$^{-1}$ and a corrected amplitude of $11.88 \pm 0.06$~mmag are obtained, which would increase the amplitude of $\nu_{\rm mod}$ by $0.18$~mmag in 2007. Therefore, $\nu_{\rm mod}$ may also have been exhibiting frequency modulation since 2007 along with the amplitude modulation shown in Fig.~\ref{figure: max plots}. 

The result of the nonlinear least-squares fit of the 2009 WASP data is within the errors of the frequency obtained from \Kepler data and thus is compatible with the fixed frequency that we use. The difference of 0.0018~d$^{-1}$ between the frequencies obtained from the linear and nonlinear fits of the 2007 WASP data is smaller than our resolution limit, and so we conclude that the linear least-squares fit using $\nu_{\rm mod}=13.3942$~d$^{-1}$ is the correct approach. The linear least-squares fit allows us to study the amplitude modulation in KIC~7106205, but we cannot explore the frequency modulation due to the poor frequency resolution.


\section{Conclusions}

This work highlights the success of utilising ground-based photometry, specifically WASP, to support \Kepler observations. We have calibrated and combined the data sets and have increased the total length of the observations of KIC~7106205 and its pulsational amplitude modulation to 6~yr. A single p~mode was extracted from the WASP observations, specifically $\nu_{\rm mod}= 13.3942$~d$^{-1}$, and has been shown to decrease in amplitude from $11.70 \pm 0.05$~mmag in 2007, to $5.87 \pm 0.03$~mmag in 2009, and to $0.58 \pm 0.06$~mmag in 2013. We agree with the conclusions of \cite{Bowman2014}, that the observed amplitude modulation in the single pulsation frequency can be explained by either a loss of mode energy to a damping region within the star; or to either invisible high-$\ell$ p~modes or two {\it internal} g~modes facilitated by the parametric resonance instability.

We are unable to extend the study of amplitude modulation in KIC~7106205 prior to 2007 using WASP data, as there are no previous observations of the star in the data archive. We have demonstrated that the amplitude of $\nu_{\rm mod}$ after 2009 is too small to be detected above the typical WASP noise level and moreover, there are no observations of KIC~7106205 after 2010. 

The methodology we have developed for combining WASP and \Kepler data can also be used to study other stars, provided that the time series overlap at some point thus allowing for the passband and dilution corrections to be made. It requires time and careful analysis to find stars that exhibit amplitude modulation, but also if they are present in the WASP data set.

The majority of the significant decrease in mode amplitude of a single pulsation frequency in KIC~7106205 occurred prior to the launch of the \Kepler mission. Therefore, it is important to remember that the \Kepler observations provide an extremely high quality, yet small {\it snapshot} of 4~yr of data, which is a mere blink of an eye insight into the stellar physics at work within this star. Time spans on the order of a decade may be more significant in \dsct stars than was previously thought and analyses of these so-called {\it coherent} pulsators should take this into account. Modelling studies are needed if we are to understand the mechanisms that cause such strong changes in pulsation mode amplitude on time scales of several years.


\section*{acknowledgements}
DMB and DLH wish to thank the STFC for the financial support of their Ph.D.s, and all authors wish to thank the \Kepler and WASP teams for providing us with such excellent data. The WASP project is funded and operated by Queen's University Belfast, the Universities of Keele, St. Andrews and Leicester, the Open University, the Isaac Newton Group, the Instituto de Astrofísica de Canarias, the South African Astronomical Observatory and by STFC. The authors wish to thank the anonymous referee for useful comments.


\bibliography{KIC7106205_WK_arxiv}


\end{document}